\documentclass[a4paper,12pt]{article}
\usepackage{amsmath,amsfonts,amssymb,cite}
\usepackage{bm}
\usepackage{relsize}

\begin{document}

\begin{center}
{\LARGE\bf A multiquark approach to excited hadrons and Regge trajectories}
\end{center}

\bigskip

\begin{center}
\renewcommand*{\thefootnote}{\alph{footnote}}
{\large S. S. Afonin
\footnote{Email: \texttt{s.afonin@spbu.ru}.}
\setcounter{footnote}{0}
}
\end{center}

\begin{center}
{\it Saint Petersburg State University, 7/9 Universitetskaya nab.,
St.Petersburg, 199034, Russia
}
\end{center}


\begin{abstract}
We propose a novel approach to construction of hadron spectroscopy.
The case of light nonstrange mesons is considered. By assumption, all such
mesons above 1~GeV appear due to creation of
constituent quark-antiquark pairs inside the $\pi$ or $\rho$($\omega$)
mesons. These spin-singlet or triplet pairs dictate the quantum
numbers of formed resonance. The resulting classification of light mesons
turns out to be in a better agreement with the experimental observations than the standard
quark model classification. It is argued that the total energy of quark components 
should be proportional to the hadron mass squared rather than the linear mass.
As a byproduct a certain relation expressing the constituent quark mass
via the gluon and quark condensate is put forward.
We show that our approach leads to an
effective mass counting scheme for meson spectrum and results in
the linear Regge and radial Regge trajectories by construction.
An experimental observation of these trajectories
might thus serve as an evidence not for string but for multiquark structure
of highly excited hadrons.
\end{abstract}


\section{Introduction}


The regularities in the masses of atomic nuclei tell us that
nuclei certainly consists of some building blocks, the nucleons,
and the mass of nuclei is just proportional to the number of nucleons.
The masses squared of light mesons also reveal many regularities~\cite{phen,phen2a}
but they are usually attributed to the properties of strong interactions
acting between valent quarks. In the present paper, we will try to
develop an alternative point of view on the masses of light hadrons ---
within the presented scheme, they could be also viewed as originating
from some effective "building blocks".

Historically the main basic framework for description of hadrons is the quark model.
This model had a great success in making order in numerous hadronic
zoo. In particular, all mesons within this picture are composed of
a quark and antiquark having the relative angular momentum $L$ which
dictates the spatial and charge parities for the
quark-antiquark systems, $P=(-1)^{L+1}$ and $C=(-1)^{L+S}$, where
$S$ is the total quark spin. By now we know that the ensuing
classification of mesons has many phenomenological flaws (see, e.g.,
the corresponding reviews in the Particle Data~\cite{pdg}). For instance,
it does not explain the scalar mesons below 1~GeV for which
a tetraquark structure is often
assumed~\cite{pelaez}. A question arises why do not we see many
light multiquark hadrons above 1~GeV? Among other questions
one can mention the experimental absence of many predicted states
(e.g., with $J^{PC}=2^{--}$) and, on the other hand, observation of
many not predicted states (e.g., too rich scalar sector, exotic $\pi_1$, etc.).

There are some deep theoretical questions as well. The light hadrons represent
highly relativistic quantum systems in which $L$ is not a
conserved quantity.
The quark angular momentum $L$ is nevertheless a standard
ingredient in constructing dynamical models for light hadrons.
Another question concerns the observable quantities like hadron
masses --- they must be renorminvariant in the field-theoretical
sense. Stated simply, they must represent some constants
independent of energy scale. The quark masses, for instance, are
not such constants since they have anomalous dimension described by QCD. A
relation of calculated observables to the renormalization
invariance is obscure in many known phenomenological models of
hadrons.

The quark model represents a concept rather than a model. A real model
appears only when a definite interaction between quarks is postulated.
Any such model should be able to explain general features of observed
hadron spectrum. In the light hadrons, perhaps the most spectacular
general feature is the observation of approximately linear Regge and
radial trajectories of the kind
\begin{equation}
\label{1b}
m_{J,n}^2=aJ+a'n+b,\qquad J,n=0,1,2,\dots.
\end{equation}
Here $J$ denotes the spin, $n$ is the radial quantum number
(enumerating the daughter Regge trajectories), $a$, $a'$ and $b$
are the slope and intercept parameters. There is no general agreement
on the position of different states on the trajectories~\eqref{1b}
(see, e.g., Refs.~\cite{phen,phen2a,phen2,2phen2,3phen2,4phen2,5phen2,6phen2,7phen2,phen3a,2phen3a,3phen3a,4phen3a,5phen3a,6phen3a,7phen3a,8phen3a,9phen3a}) but the existence of these trajectories seems
to be certain for many phenomenologists~\cite{phen3a,2phen3a,3phen3a,4phen3a,5phen3a,6phen3a,7phen3a,8phen3a,9phen3a,phen3,2phen3,3phen3,4phen3,5phen3,6phen3,7phen3}. There are also some indications
on a similar behavior in the heavy quark sector~\cite{phen4,2phen4,3phen4,4phen4,5phen4}. The non-relativistic
and partially relativistic potential models cannot explain the
recurrences~\eqref{1b} in a natural way. Usually the observation of linear
trajectories~\eqref{1b} is interpreted as an evidence for string picture
of mesons. In spite of many attempts (see, e.g., Refs.~\cite{string,string2,string3,string4,string5,string6,string7,string8}), however,
a satisfactory quantized hadron string has not been constructed.
Among typical flaws of this approach one can mention the absence of
spontaneous chiral symmetry breaking, rapidly growing (with mass)
size of meson excitations that is not supported experimentally,
unclear role of higher Fock components in the hadron wavefunction.

The purpose of the present work is to propose a novel realization
of quark model concept, a realization that leads to a natural
(and alternative to hadron strings) explanation of Regge recurrences~\eqref{1b}
and that is potentially free of typical shortcomings inherent to string,
potential and some other approaches.

\section{Masses and classification of light mesons}

In hadron physics, the pion is known to be the most important and best studied
meson both experimentally and theoretically. Concerning the theoretical aspect,
the pion is the only hadron (along with $K$ and $\eta$) for which we know a
model-independent mass formula --- the Gell-Mann--Oakes--Renner (GOR) relation~\cite{gor},
\begin{equation}
\label{1}
m_\pi^2=-\frac{\langle\bar{q}q\rangle}{f_\pi^2}(m_u+m_d)=\Lambda\cdot2m_q,
\end{equation}
where we set $m_u=m_d\doteq m_q$ and
$\Lambda\doteq-\frac{\langle\bar{q}q\rangle}{f_\pi^2}$. We will
use the standard values for the quark condensate and masses of
current quarks at the scale of the pion mass from the QCD sum
rules~\cite{svz,svz2} and Chiral Perturbation Theory
(ChPT)~\cite{gasser},
$\langle\bar{q}q\rangle=-(250\,\text{MeV})^3$, $m_u+m_d=11$~MeV.
Using $f_\pi=92.4$~MeV (the value of weak pion decay
constant
), the
relation~\eqref{1} yields $m_\pi=140$~MeV and $\Lambda=1830$~MeV.
The famous relation~\eqref{1} was derived in various approaches
assuming the spontaneous
Chiral Symmetry Breaking (CSB) in strong interactions.

The renormalization invariance of pion mass follows from the
renorminvariance of operator $m_q\bar{q}q$ in QCD Lagrangian. Masses
of other mesons must be also renorminvariant. On the other hand,
the light nonstrange mesons can be viewed, in one way or another, as
some quantum excitations of pion. We will assume that the masses of these
mesons can be represented as
\begin{equation}
\label{2b}
m_h^2=\Lambda (E_h+2m_q)=\Lambda E_h+m_\pi^2.
\end{equation}
Below we will motivate a proportionality of the energy parameter $E_h$
to the renorminvariant gluon condensate, $E_h\sim\alpha_s\langle
G_{\mu\nu}^2\rangle/\langle\bar{q}q\rangle$. But first we would like to
develop a phenomenological mass counting scheme based on the relation~\eqref{2b}
and demonstrate how it describes the meson spectroscopy.

Our second assumption is that the parameter $E_h$ can be interpreted as an
effective energy of some constituents different from the current quarks.
The problem is to propose a model for these constituents which should
represent some excitations inside the pion. We postulate three basic
excitations. The first one appears when one of quarks absorbs a gluon
of certain energy $E_h=E_\rho$. The spin of excited quark changes its
direction to the opposite one, converting the original spin-singlet
$q\bar{q}$-pair ($\pi$-meson) to the spin-triplet one ($\rho$-meson).
The second kind of excitation emerges due to formation of
spin-singlet $q\bar{q}$-pair with effective mass $E_h=E_0$ (the lower
index stays for total spin of $S$-wave $q\bar{q}$-pair). We will call
this pair $Q_0$. The third basic
excitation is the formation of
spin-triplet $q\bar{q}$-pair with effective mass $E_h=E_1$. This pair
will be called $Q_1$. The formation of constituent pairs $Q_0$ and $Q_1$
may be a QCD
analogue of formation of para- and ortho-positronia from photons.
We assume that any excitation inside pion leading to an observable
resonance can be represented as a combination of these
basic excitations so that the total effective energy $E_h$ in~\eqref{2b}
is just a sum (appropriate number of times) of $E_\rho$, $E_0$, and $E_1$.
Simultaneously this will dictate the quantum numbers of constructed resonance
($Q_0$ and $Q_1$ possess $J^{PC}=0^{-+}$ and $1^{--}$, respectively).
In a sense, the introduction of $Q_0$ and $Q_1$ is our model for excitations
of higher Fock components in the pion wavefunction.

It is convenient to divide the nonstrange meson resonances above 1~GeV into
excited pions and excited $\rho$($\omega$) mesons. The excited
$\rho$ mesons containing $n$ pairs $Q_0$ and $l$
pairs $Q_1$ will be labeled as
$^J\!{\mathlarger{\mathlarger{\rho}}}_{{\!Q_0^n Q_1^l}}$, where $J$ means
the total spin. According to our mass counting scheme, masses of
these excitations are given by the relation
\begin{equation}
\label{5}
m_{{\mathlarger{\mathlarger{\rho}}}_{{\!Q_0^n Q_1^l}}}^2=\Lambda(E_\rho+nE_0+lE_1)+m_\pi^2.
\end{equation}
By assumption, the whole system is in the
$S$-wave state\footnote{One can include into this scheme the higher $L>0$ waves,
we will consider the simplest possibility.}, hence,
at fixed $n$ and $l$ the relation~\eqref{5} defines mass
of a set of degenerate states with various spins up to $J=l+1$.
The $P$ and $C$ parities obey to the standard multiplicative law:
$(P,C)=((-1)^{n+l+1},(-1)^{l+1})$. The total isospin of system of
$n+l$ constituent $q\bar{q}$ pairs is zero as this system, by assumption,
arises from an isosinglet combination of gluons.

The excited pions follow the
same principle: The states
$^J{\mathlarger{\mathlarger{\pi}}}_{{\!\! Q_0^n Q_1^l}}$ have mass
\begin{equation}
\label{5b}
m_{{\mathlarger{\mathlarger{\pi}}}_{{\!\!Q_0^n Q_1^l}}}^2=\Lambda(nE_0+lE_1)+m_\pi^2,
\end{equation}
maximal spin $J=l$, and parities $(P,C)=((-1)^{n+l+1},(-1)^l)$.

Since a simultaneous excitation of two close pairs $Q_0Q_0$ does not change
neither spin nor parities, it is natural to refer to
a meson containing even number $2k$ of $Q_0$ and no one $Q_1$
as the $k$-th radial excitation.
The radial excitations appear thus with a
"period" $2E_0\Lambda$.

Consider a formation of two close vector pairs $Q_1 Q_1$. As they are
formed from originally massless gluons we must combine their spins
as if they were massless particles. This follows from conservation of
total spin of original gluons. We recall that the spin of massless
gluons has only two projections--- two possible helicities. The
total helicity of two gluons can be 0 or 2.
This rule\footnote{In other words, the
addition law for spins of spin-1 massless particles is the same as
for massive spin-$\frac12$ particles, just the final result must
be multiplied by 2.} extends to arbitrary number of close $Q_1$ pairs.
It is easy to see now that the excitations of the kind
$^J\!{\mathlarger{\mathlarger{\rho}}}_{{\!Q_1^{2k}}}$,
$k=0,1,2,\dots$, will give rise to a degenerate family of
resonances with $J^{PC}=(1,3,\dots,2k+1)^{--}$. The states on the
main $\rho$-meson Regge trajectory follow thus with a "period"
$2E_1\Lambda$. These excitations generate also daughter
trajectories. For instance, the additional $\rho$ mesons ($J^{PC}=1^{--}$) arise
which appear with the same "period" $2E_1\Lambda$.
They are different from the "radial" $\rho$ mesons that appeared with
another period $2E_0\Lambda$.

In order to demonstrate how our scheme works in practice
let us consider some examples.

The mass of $\rho$-meson is $m_\rho=\sqrt{\Lambda E_\rho+m_\pi^2}$. This
relation fixes $E_\rho\approx310$~MeV from the
averaged mass of charged $\rho$ in the hadronic processes,
$m_\rho=766.5\pm1.1$~MeV~\cite{pdg}.

$^1\!{\mathlarger{\mathlarger{\rho}}}_{{\!Q_0}}$ and
$^1{\mathlarger{\mathlarger{\omega}}}_{{\!Q_0}}$ have quantum
numbers $J^{PC}=1^{+-}$ and, according to Eq.~\eqref{5}, mass
$m=\sqrt{\Lambda (E_\rho+E_0)+m_\pi^2}$. Fitting to the
masses of experimental states $h_1(1170)$ and $b_1(1230)$~\cite{pdg},
we obtain the estimate $E_0\approx430-510$~MeV.

The state $^0{\mathlarger{\mathlarger{\pi}}}_{{\!\! Q_0}}$ has the
quantum numbers of scalar particle, $J^{PC}=0^{++}$.
According to Eq.~\eqref{5b} its mass is
$m=\sqrt{\Lambda E_0+m_\pi^2}\approx900-980$~MeV. This estimate
is close to the mass of $a_0(980)$~\cite{pdg}.

$^J\!{\mathlarger{\mathlarger{\rho}}}_{{\!Q_1}}$ and
$^J{\mathlarger{\mathlarger{\omega}}}_{{\!Q_1}}$ have
$J^{PC}=(0,1,2)^{++}$. They should be the series of states
$a_1(1260)$, $f_1(1285)$, $f_2(1270)$, $a_2(1320)$, and likely
$f_0(1370)$~\cite{pdg}. Substituting the experimental masses of well measured
resonances $f_1(1285)$ and $f_2(1270)$ to our relation for their mass,
$m=\sqrt{\Lambda (E_\rho+E_1)+m_\pi^2}$, we obtain the estimate
$E_1\approx570-580$~MeV.

$^1\!{\mathlarger{\mathlarger{\rho}}}_{{\!Q_0^2}}$ and
$^1{\mathlarger{\mathlarger{\omega}}}_{{\!Q_0^2}}$ have mass
$m=\sqrt{\Lambda (E_\rho+2E_0)+m_\pi^2}\approx1470-1590$~MeV.
They are the first radial excitations of $\rho$ and $\omega$
mesons and describe the resonance regions $\rho(1450)$ and
$\omega(1420)$~\cite{pdg}.

$^J\!{\mathlarger{\mathlarger{\rho}}}_{{\!Q_0Q_1}}$ and
$^J{\mathlarger{\mathlarger{\omega}}}_{{\!Q_0Q_1}}$ have quantum
numbers $J^{PC}=(0,1,2)^{-+}$, i.e. 3 possible spins and pion
parities. These states describe resonances $\pi$, $\pi_1$ and
$\pi_2$ above 1~GeV. In contrast to the standard quark model, the
$\pi_1$-meson is not exotic in our scheme! The relation~\eqref{5b}
predicts mass about
$m=\sqrt{\Lambda (E_\rho+E_0+E_1)+m_\pi^2}\approx1550-1610$~MeV.
The possible candidates are
$\pi_1(1600)$ and $\pi_2(1670)$~\cite{pdg}.

The states $^J\!{\mathlarger{\mathlarger{\rho}}}_{{\!Q_1^2}}$ and
$^J{\mathlarger{\mathlarger{\omega}}}_{{\!Q_1^2}}$ have
$J^{PC}=(1,3)^{--}$ as the total spin of the pair $Q_1Q_1$
can be 0 or 2. Natural candidates are
$\rho(1700)$ and $\rho_3(1690)$ (and $\omega(1650)$ with
$\omega_3(1670)$ for $\omega$)~\cite{pdg}.
Our fits yield $m=\sqrt{\Lambda (E_\rho+2E_1)+m_\pi^2}\approx1630-1650$~MeV.
As in the standard quark model, the exotic states
with quantum numbers $J^{PC}=0^{--}$ are absent and
$\rho(1700)$ is not the second radial excitation\footnote{In the
potential quark models, $\rho(1700)$ (and $\omega(1650)$) is a
$D$-wave state, while the radial excitations of $\rho(770)$ should
be $S$-wave ones.} of $\rho(770)$. On the other hand, we do not obtain
states with $J^{PC}=2^{--}$ which are predicted by the standard quark model
but were not observed.

\section{Regge trajectories}

The relations~\eqref{5} and~\eqref{5b} lead to linear Regge,
equidistant daughter Regge and radial trajectories. Below we give
examples for some of them.

The states $^0{\mathlarger{\mathlarger{\pi}}}_{{\!\! Q_0^{2n}}}$,
$n=0,1,2,\dots$, form linear trajectory for the radial excitations
of pion,
\begin{equation}
\label{6}
m_\pi^2(n)=2\Lambda E_0n+m_\pi^2.
\end{equation}
The radial excitations of $\rho(770)$ --- the resonances
$^1\!{\mathlarger{\mathlarger{\rho}}}_{{\!Q_0^{2n}}}$ --- lie on
the first radial $\rho$-trajectory,
\begin{equation}
\label{7}
m_\rho^2(n)_I=2\Lambda E_0\left(n+\frac{E_\rho}{2E_0}\right)+m_\pi^2.
\end{equation}
The second radial $\rho$-trajectory is composed of the states
$^1\!{\mathlarger{\mathlarger{\rho}}}_{{\!Q_1^2Q_0^{2n}}}$,
\begin{equation}
\label{8}
m_\rho^2(n)_{I\!I}=2\Lambda E_0\left(n+\frac{E_1}{E_0}+\frac{E_\rho}{2E_0}\right)+m_\pi^2.
\end{equation}
It is evident that the states
$^1\!{\mathlarger{\mathlarger{\rho}}}_{{\!Q_1^{2(\!k-1\!)}Q_0^{2n}}}$
formally give rise to the $k$-th radial $\rho$-trajectory. The
resonances having structure
$^1\!{\mathlarger{\mathlarger{\rho}}}_{{\! Q_1Q_0^{2n}}}$ form the
first radial $a_1$-trajectory,
\begin{equation}
\label{9}
m_{a_1}^2(n)_I=2\Lambda E_0\left(n+\frac{E_1}{2E_0}+\frac{E_\rho}{2E_0}\right)+m_\pi^2.
\end{equation}
The expressions for further axial radial trajectories can be easily written.
It should be remarked that the standard quark model predicts only
two radial $\rho$-trajectories ($S$- and $D$-wave ones) and one
$a_1$-trajectory (a $P$-wave one). The scheme under discussion is much richer.

The spin $\rho$-trajectory is composed of the states
$^J\!{\mathlarger{\mathlarger{\rho}}}_{{\!Q_1^{J-1}}}$ with
$J=1,3,5,\dots$. The corresponding masses are
\begin{equation}
\label{10}
m_{\rho_J}^2=\Lambda E_1\left(J-1+\frac{E_\rho}{E_1}\right)+m_\pi^2.
\end{equation}
For the even spins, $J=2,4,6,\dots$, the trajectory~\eqref{10}
describes $a_J$ ($f_J$) mesons. The Regge trajectory~\eqref{10}
describes thus states with alternating parities and this agrees
with the phenomenology.


The principal $\rho$-meson Regge trajectory~\eqref{10} is
accompanied by the daughter trajectories following with the step
$2\Lambda E_1$. The spin-1 $\rho$-mesons of the kind
$^1\!{\mathlarger{\mathlarger{\rho}}}_{{\!Q_1^{J-1}}}$ are the
lowest states on the daughters. For example, the lowest state
in~\eqref{8} is the lowest state on the first daughter. The
spectrum of $^1\!{\mathlarger{\mathlarger{\rho}}}_{{\!Q_1^{J-1}}}$
excitations reads
\begin{equation}
\label{10b}
m_{^1\!{\mathlarger{\mathlarger{\rho}}}_{{\!Q_1^{J-1}}}}^2=
2\Lambda E_1\left(k+\frac{E_\rho}{2E_1}\right)+m_\pi^2,
\end{equation}
where $k=0,1,2,\dots$ enumerates the daughters. Similar relations
can be written for the axial and other mesons. An obvious
consequence of the emerging spectrum is the degeneracy of spin and
daughter radial excitations of the type $m^2(J,k)\sim J+k$, which
is typical for the Veneziano dual amplitudes~\cite{avw} and the
Nambu--Goto open strings. This kind of degeneracy was observed in the
experimental spectrum of light nonstrange mesons~\cite{phen2,2phen2,3phen2,4phen2,5phen2,6phen2,7phen2}.

The fitted values of $E_\rho$, $E_0$ and $E_1$ are rather close.
This allows to consider reasonable limits where some of them are equal.
The notions of radial and daughter radial trajectories coincide in
the limit $E_0=E_1$. In the limit $E_\rho=E_1$, the radial vector
and axial trajectories are related by
\begin{equation}
\label{11}
m_{a_1}^2(n)=m_\rho^2(n)+m_\rho^2-m_\pi^2.
\end{equation}
This relation holds both for radial and for daughter radial
trajectories. In the chiral limit, $m_\pi=0$, the
relation~\eqref{11} for the ground states, $n=0$, reduces to the
old Weinberg relation, $m_{a_1}^2=2m_\rho^2$~\cite{wein}. In the
most symmetric limit, $E_\rho=E_0=E_1$, the vector and axial
radial spectrum in the chiral limit reduce to a very simple form,
\begin{equation}
\label{12}
m_\rho^2(n)=2m_\rho^2\left(n+\frac12\right),
\quad
m_{a_1}^2(n)=2m_\rho^2\left(n+1\right).
\end{equation}
The relations~\eqref{12} first appeared in the
variants of Veneziano amplitude which incorporated the Adler
self-consistency condition~\cite{avw}.
This condition (the amplitude of $\pi\pi$ scattering is zero at zero momentum)
incorporates the CSB removing degeneracy between the $\rho$ and $a_1$ spectra .
Within the QCD sum rules, the relations~\eqref{12} may be
interpreted as a large-$N_c$ generalization of the Weinberg
relation~\cite{afonin:plb}.

We see thus that in certain limits the Regge
phenomenology of our approach reproduces various known relations.
In addition, it is not excluded that the assumption $E_\rho=E_0=E_1$
is compatible with the available data if a global fit is performed.
According  to the phenomenological analysis in review~\cite{phen2a}
the averaged slopes of spin and radial trajectories in light nonstrange
mesons, $a$ and $a'$ in Eq.~\eqref{1b}, are equal; the reported value is
$a\approx a'\approx1.14$~GeV$^2$. This remarkable observation leads
to a large degeneracy in the spectrum~\cite{phen2,2phen2,3phen2,4phen2,5phen2,6phen2,7phen2,phen3,2phen3,3phen3,4phen3,5phen3,6phen3,7phen3}. From the
given value we have
$E_0\approx E_1\approx a/(2\Lambda)\approx1.14/(2\cdot1.83)\approx0.31$~GeV
that coincides with our previous estimate for $E_\rho$.

Let us clarify further how the excited resonances with identical
quantum numbers can have different origin in the proposed scheme.
They may represent the radial states, states on daughter Regge
trajectories and various "mixed" ones. For instance, the second
$\rho$-meson excitation with the same parities
appears in three forms: the second radial
excitation $^1\!{\mathlarger{\mathlarger{\rho}}}_{{\!Q_0^4}}$, the
vector state on the second daughter trajectory
$^1\!{\mathlarger{\mathlarger{\rho}}}_{{\!Q_1^4}}$, and the mixed
one $^1\!{\mathlarger{\mathlarger{\rho}}}_{{\!Q_0^2Q_1^2}}$. They
are degenerate only in the limit $E_0=E_1$. It is likely difficult
to detect such a splitting experimentally because of overlapping
widths. In reality, one would observe rather a "broad resonance
region".

We note also that placing of the observed "radial" states on a
certain trajectory should be made with care --- an incorrect
interpretation of states leads to a false (or more precisely,
introduced by hands) non-linearity of the trajectory. Take again
the $\rho$ meson as a typical example. The first three radial
excitations of $\rho$ are the states
$^1\!{\mathlarger{\mathlarger{\rho}}}_{{\!Q_0^2}}$,
$^1\!{\mathlarger{\mathlarger{\rho}}}_{{\!Q_0^4}}$, and
$^1\!{\mathlarger{\mathlarger{\rho}}}_{{\!Q_0^6}}$. They are
accompanied by the following states with the quantum numbers of
$\rho$: $^1\!{\mathlarger{\mathlarger{\rho}}}_{{\!Q_1^2}}$,
$^1\!{\mathlarger{\mathlarger{\rho}}}_{{\!Q_1^2Q_0^2}}$, and
$^1\!{\mathlarger{\mathlarger{\rho}}}_{{\!Q_1^4}}$. Since $\frac32
E_0\!>\!E_1\!>\!E_0$ in our fits, the sequence of first 7
$\rho$-mesons is: $\mathlarger{\mathlarger{\rho}}$,
$^1\!{\mathlarger{\mathlarger{\rho}}}_{{\!Q_0^2}}$,
$^1\!{\mathlarger{\mathlarger{\rho}}}_{{\!Q_1^2}}$,
$^1\!{\mathlarger{\mathlarger{\rho}}}_{{\!Q_0^4}}$,
$^1\!{\mathlarger{\mathlarger{\rho}}}_{{\!Q_1^2Q_0^2}}$,
$^1\!{\mathlarger{\mathlarger{\rho}}}_{{\!Q_1^4}}$,
$^1\!{\mathlarger{\mathlarger{\rho}}}_{{\!Q_0^6}}$. They likely
correspond to the vector resonances $\rho(770)$, $\rho(1450)$,
$\rho(1700)$, $\rho(1900)$, $\rho(2000)$, $\rho(2150)$,
$\rho(2270)$~\cite{pdg}.

\section{A theoretical motivation}

It is interesting to get a theoretical hint both on the rule
$m_h^2\sim E_{\rho,0,1}$ underlying the presented mass counting scheme
and on the value of energy parameters $E_{\rho,0,1}$. The latter point
is intriguing by itself since the scale about $0.3$~GeV is ubiquitous
in the hadron physics, for instance it often emerges
in mass relations between hadrons containing heavy quarks~\cite{Afonin:2014tda}.

The mass of a hadron state $|h\rangle$ can be related to the trace
of energy momentum tensor $\Theta_{\mu\nu}$ in QCD
(the sign convention is mostly positive),
\begin{equation}
\label{2}
2m_h^2=-\langle h|\Theta^\mu_\mu|h\rangle,
\end{equation}
where $\Theta^\mu_\mu$ is given by the scale anomaly,
\begin{equation}
\label{3}
\Theta^\mu_\mu=\frac{\beta}{2g_s}G_{\mu\nu}^2+\!\!\sum_{q=u,d,\dots}\!\!\!\! m_q\bar{q}q.
\end{equation}
Here $\beta$ denotes the QCD Beta-function and $g_s$ is the coupling constant.
The relation~\eqref{2} represents trace taken in the
following Ward identity
known in deep inelastic phenomenology~\cite{jaffe}:
$2p_\mu p_\nu=-\langle h(p_\mu)|\Theta_{\mu\nu}|h(p_\nu)\rangle$.
As follows from
derivation of the identity~\eqref{2}, the squared mass (not the
linear one!) in the l.h.s. appears due to the relativistic
invariance.
The r.h.s. of Eq.~\eqref{2} is a renorminvariant quantity. One
can build two substantially different renormalization invariant
operators in QCD, both these operators are present in~\eqref{3}.
The l.h.s. of Eq.~\eqref{2} defines the gravitational mass of a hadron.
In reality, the vast majority of hadron resonances probably have no
well defined gravitational (and inertial) mass as they do not
propagate in space: Their typical lifetime of the order of
$10^{-23}$--$10^{-24}$s does not allow to leave the reaction area
of the order of 1~fm which is comparable to their size. They show
up only as some structures in the physical observables at certain
energy intervals. The resonance mass is commonly associated with
the real part of an $S$-matrix pole on the second (unphysical)
sheet. How this definition is related with the gravitational mass
is by far not obvious in a theory with confinement.

In the sector of light $u$ and $d$ quarks, there are only two hadrons
with well defined gravitational mass --- the pion and nucleon. The
pion mass is given by the GOR relation~\eqref{1}. This relation should
somehow follow from the Ward identity~\eqref{2}. An explicit derivation
of~\eqref{1} from~\eqref{2} would likely give an analytical proof for
spontaneous CSB in QCD. Our present aim is to show,
at least on a rather intuitive level, how the mass gap may emerge in
non-perturbative gluon vacuum and get some quantitative insight.
Consider the case of nucleon,
$|h\rangle=|N\rangle$, for which the Eq.~\eqref{2} must yield the
proton mass. Let us insert a complete set of vacuum states
$|0\rangle \langle0|$ from both sides of $\Theta^\mu_\mu$ in Eq.~\eqref{2},
the nucleon mass is then given by
\begin{equation}
\label{3b}
m_N^2=-\frac12\langle 0|N\rangle^2\langle 0| \frac{\beta}{2g_s}G_{\mu\nu}^2+\!\!\sum_{q=u,d}\!\! m_q\bar{q}q|0\rangle.
\end{equation}
This expression suggests that in the chiral limit, $m_q=0$, the nucleon mass is
determined by the vacuum average $\langle 0|G_{\mu\nu}^2|0\rangle$.

From the one-loop QCD beta-function we have
$\frac{\beta}{2g_s}=-\frac{\beta_0}{8}\frac{\alpha_s}{\pi}$, where
$\beta_0=11-\frac23n_f$, $\alpha_s=\frac{g_s^2}{4\pi}$. The relation~\eqref{3b}
can be then rewritten as
\begin{equation}
\label{3c}
m_N^2=\frac12\langle 0|N\rangle^2\langle\bar{q}q\rangle\left(-\sum_{q} m_q + \frac{\beta_0}{8}\frac{\frac{\alpha_s}{\pi}\langle
G_{\mu\nu}^2\rangle}{\langle\bar{q}q\rangle} \right).
\end{equation}
The relation~\eqref{3c} demonstrates that the masses of current quarks in nucleon
acquire a contribution from non-perturbative gluon vacuum. Since nucleon
at rest is interpreted in the quark model as a bound system of
three constituent quarks, we can estimate from the Eq.~\eqref{3c} an
effective energy per quark, i.e. the value of constituent quark mass ---
in the chiral limit, it is just one third of gluon
contribution in~\eqref{3c},
\begin{equation}
\label{3d}
M_c\simeq-\frac{\beta_0}{24}\frac{\frac{\alpha_s}{\pi}\langle
G_{\mu\nu}^2\rangle}{\langle\bar{q}q\rangle}.
\end{equation}
Substituting into~\eqref{3d} the standard value of gluon condensate
from QCD sum rules, $\frac{\alpha_s}{\pi}\langle
G_{\mu\nu}^2\rangle=0.012(3)$~GeV$^4$~\cite{svz,svz2} and
$\langle\bar{q}q\rangle=-(0.25\,\text{GeV})^3$, $n_f=2$
we obtain the estimate $M_c\simeq310$~MeV. It is a typical estimate for the value
of constituent quark mass\footnote{Strictly speaking, the value of constituent quark
mass is very model dependent --- as far as we know, it ranges from
220 to 450~MeV in various models. The value
$M_c(p^2)\simeq310$~MeV at small momentum $p$, however, proves to
be seen in unquenched lattice simulations in the chiral limit
(see, e.g., Ref.~\cite{bowman}). We can indicate a simple qualitative way
leading to this estimate. When one incorporates the vector and
axial mesons into the low-energy models describing the sponataneous CSB in QCD
and related physics, e.g. into the Nambu--Jona-Lasinio
model~\cite{klev} or the linear sigma-model, a model-independent
relation emerges: $m_{a_1}^2=m_\rho^2+6M_c^2$. On the other hand,
the idea of CSB was also exploited in the derivation of famous
Weinberg relation, $m_{a_1}^2=2m_\rho^2$~\cite{wein}. Combining
both relations, one has $M_c=m_\rho/\sqrt{6}$. Substitution of
the value $m_\rho\approx766$~MeV used in our fits leads to
$M_c\simeq0.31$~GeV.}, $M_c\approx m_p/3$, where $m_p$ is the
proton mass.

It is interesting to observe the numerical coincidence $M_c\simeq E_\rho$.
This suggests that the "spin flip" converting pion to $\rho$ is
essentially equivalent to creation of constituent quark.
Thus the $\rho$ meson in our approach may be interpreted as
a system of one constituent quark and one current antiquark
(or vice versa) which interact with the QCD vacuum.
This picture is different from the usual quark model where $\rho$
consists of two constituent quarks interacting with each
other\footnote{In view of these speculations one can ask what
is the exact relation for the nucleon mass within the given approach?
We can propose the following guess. In the case of $\rho$ meson, we motivated
"a rule of transition" from the non-relativistic quark model to our scheme:
$m_h=\sum M_c+\text{interactions}$ $\rightarrow$ $m_h^2=\Lambda\cdot\frac12\sum M_c+m_\pi^2$.
Applying this "rule" to the ground state of nucleon (3 constituent quarks)
we arrive at the relation $m_N^2=\frac32\Lambda M_c+m_\pi^2$ that gives $m_N\simeq0.93$~GeV.
The factor $\frac32$ could be also qualitatively interpreted as follows:
The creation of 3 constituent quarks and 3 constituent antiquarks (i.e., the creation of
$N\bar{N}$ pair in vacuum) is equivalent to creation of 3 constituent quark-antiquark pairs,
so the relation of energies of constituents in nucleon and vector meson is 3 to 2.}.

The given interpretation can be further substantiated by the
observation that the same "spin flip" should convert the nucleon
to $\Delta$ baryon. As this flip "costs" $\Lambda M_c$ for the hadron
mass squared, we should expect the relation $m^2_\Delta=m^2_N+\Lambda M_c$.
It indeed yields $m_\Delta\simeq1.2$~GeV in a good agreement with
experimental data for $\Delta(1232)$~\cite{pdg}.

Coming back to motivation of our approach, the presented mass
counting scheme is based on the assumption that the origin of
hadron masses is similar to the case of nucleon mass in
Eq.~\eqref{3c} --- the hadron mass squared (not the linear one
as in many other approaches!) is proportional to
effective energy of hadron constituents. This rule is
conjectured from the Ward identity~\eqref{2}.

The second assumption refers to the postulated form of meson
constituents --- the valent $q\bar{q}$ pair plus constituent
$q\bar{q}$ pairs which effectively parametrize contributions
to hadron mass from strong gluon field. The given choice is a model
which works in the meson spectroscopy above and near 1~GeV. Besides the spectroscopy
it could explain why the highly excited nonstrange mesons
prefer to decay to more than two pions or, say, to have sometimes
$\rho$-meson in the decay products instead of
$\pi\pi$-pair~\cite{pdg} --- the corresponding possibilities are
likely enciphered in the expression for the mass of a resonance
as is seen in the examples of Section~2.

It should be mentioned that hadron structure strongly depends on a reference frame.
For instance, the proton structure experimentally looks very different
if it probed for a proton at rest or for ultrarelativistic proton.
It might be that the hadron model developed in this paper is more
appropriate as a picture of light hadrons near the light cone while the traditional quark models
refer to the rest frame. In this case there is no contradiction between
different approaches.

The constituent $q\bar{q}$ pairs are not operative degrees of
freedom noticeably below 1~GeV. It is interesting to note, however,
that the spectroscopy in this sector can be constructed following
essentially the same relation~\eqref{2b} if the pseudoscalar pair $Q_0$
is replaced by the pseudogoldstone bosons $\pi$, $K$, $\eta$~\cite{Afonin:2017wjj}.

\section{Summary}

We have proposed a novel approach to classification of mesons and
description of meson spectroscopy. The approach represents a new
realization of the quark model concept, a realization in which
highly excited states appear due to multiquark components in
hadron wavefunctions. We constructed a mass counting scheme that
allows to estimate meson masses by a trivial arithmetics with
accuracy comparable to numerical calculations in complicated dynamical models.
The given scheme is based on introduction of constituent scalar
and vector quark-antiquark pairs. These pairs in excited hadrons,
in some sense, bear a superficial resemblance to neutrons and protons
in atomic nuclei.

The inclusion of strange quarks into our approach is straightforward.
It would be interesting to extend the presented ideas to light
baryons and to hadrons with heavier quarks.

The proposed approach is
broader than "just another one model" as it provides a new framework
for analysis of hadron resonances
and can be used as a
starting point for construction of essentially new dynamical
models in hadron physics.

\section*{Conflicts of Interest}

The author declares that there is no conflict of interest regarding the publication of this paper.


\begin{thebibliography}{99}
\bibitem{phen} A.~V.~Anisovich, V.~V.~Anisovich and A.~V.~Sarantsev,
Phys. Rev. D~{\bf 62}, 051502(R) (2000).
\bibitem{phen2a} D.~V.~Bugg, Phys. Rept.
{\bf 397}, 257 (2004).
\bibitem{pdg} M. Tanabashi {\it et al.} (Particle Data Group), Phys. Rev. D {\bf 98}, 030001 (2018).
\bibitem{pelaez}
  J.~R.~Pelaez,
  Phys. Rept. {\bf 658}, 1 (2016).
\bibitem{phen2} E. Klempt and A. Zaitsev, Phys. Rep. {\bf 454}, 1 (2007).
\bibitem{2phen2}
M. Shifman and A. Vainshtein, Phys. Rev. D {\bf 77}, 034002 (2008).
\bibitem{3phen2}
S.~S.~Afonin, Eur. Phys. J. A {\bf 29}, 327 (2006).
\bibitem{4phen2}
S.~S.~Afonin, Phys. Lett. B {\bf 639}, 258 (2006).
\bibitem{5phen2}
S.~S.~Afonin, Mod. Phys. Lett. A {\bf 22}, 1359 (2007).
\bibitem{6phen2}
S.~S.~Afonin, Int. J. Mod. Phys. A {\bf 22}, 4537 (2007).
\bibitem{7phen2}
S.~S.~Afonin, Phys. Rev. C {\bf 76}, 015202 (2007).
\bibitem{phen3a} D.~M.~Li, B.~Ma, Y.~X.~Li, Q.~K.~Yao and H.~Yu,
Eur.\ Phys.\ J.\ C {\bf 37}, 323 (2004).
\bibitem{2phen3a}
P.~Masjuan, E.~Ruiz Arriola and W.~Broniowski,
Phys. Rev. D {\bf 85}, 094006 (2012).
\bibitem{3phen3a}
E. Klempt, Phys. Rev. C {\bf 66}, 058201 (2002).
\bibitem{4phen3a}
J.~Sonnenschein and D.~Weissman,
JHEP {\bf 1408}, 013 (2014).
\bibitem{5phen3a} J.~Sonnenschein and D.~Weissman,
JHEP {\bf 1502}, 147 (2015).
\bibitem{6phen3a} J.~Sonnenschein and D.~Weissman,
JHEP {\bf 1512}, 011 (2015).
\bibitem{7phen3a}
C.~Q.~Pang, B.~Wang, X.~Liu and T.~Matsuki,
Phys.\ Rev.\ D {\bf 92}, 014012 (2015).
\bibitem{8phen3a}
A.~M.~Badalian and B.~L.~G.~Bakker,
Phys.\ Rev.\ D {\bf 93}, 074034 (2016).
\bibitem{9phen3a}
D.~Jia, C.~Q.~Pang and A.~Hosaka,
Int.\ J.\ Mod.\ Phys.\ A {\bf 32}, 1750153 (2017).
\bibitem{phen3} P. Bicudo, Phys. Rev. D {\bf 76}, 094005 (2007).
\bibitem{2phen3} P. Bicudo, {\bf 81}, 014011 (2010).
\bibitem{3phen3} L.~Y.~Glozman,
Phys.\ Rept.\  {\bf 444}, 1 (2007).
\bibitem{4phen3} E.~H.~Mezoir and P.~Gonzalez, Phys. Rev. Lett. {\bf 101}, 232001 (2008).
\bibitem{5phen3} S.~S.~Afonin, Int. J. Mod. Phys. A {\bf 23}, 4205 (2008).
\bibitem{6phen3} S.~S.~Afonin, Mod. Phys. Lett. A {\bf 23}, 3159 (2008).
\bibitem{7phen3} T.~D.~Cohen,
Nucl.\ Phys.\ Proc.\ Suppl.\  {\bf 195}, 59 (2009).
\bibitem{phen4} S.~S.~Gershtein, A.~K.~Likhoded and
A.~V.~Luchinsky, Phys. Rev. D {\bf 74}, 016002 (2006).
\bibitem{2phen4} S.~S.~Afonin and I.~V.~Pusenkov,
Phys.\ Rev.\ D {\bf 90}, 094020 (2014).
\bibitem{3phen4} S.~S.~Afonin and I.~V.~Pusenkov, Mod.\ Phys.\ Lett.\ A {\bf 29}, 1450193 (2014).
\bibitem{4phen4} P.~Masjuan, E.~Ruiz Arriola and W.~Broniowski,
EPJ Web Conf. {\bf 73}, 04021 (2014).
\bibitem{5phen4} K.~Chen, Y.~Dong, X.~Liu, Q.~F.~Lu and T.~Matsuki,
  Eur.\ Phys.\ J.\ C {\bf 78}, 20 (2018).
\bibitem{string} J.~Sonnenschein,
  Prog.\ Part.\ Nucl.\ Phys.\  {\bf 92}, 1 (2017).
\bibitem{string2} Y. Nambu, Phys. Rev. D {\bf 10}, 4262 (1974).
\bibitem{string3} D. LaCourse and M. G. Olsson, Phys. Rev. D {\bf 39}, 2751 (1989).
\bibitem{string4} A. Yu. Dubin, A. B. Kaidalov and Yu. A. Simonov, Phys. Lett. B {\bf 323}, 41 (1994).
\bibitem{string5} Yu. S. Kalashnikova, A. V. Nefediev and Yu. A. Simonov, Phys. Rev. D {\bf 64}, 014037 (2001).
\bibitem{string6} T. J. Allen, C. Goebel, M. G. Olsson and S. Veseli Phys. Rev. D {\bf 64}, 094011 (2001).
\bibitem{string7} M. Baker and R. Steinke, Phys. Rev. D {\bf 65}, 094042 (2002).
\bibitem{string8} F. Buisseret, Phys. Rev. C {\bf 76}, 025206 (2007).
\bibitem{gor} M. GelI-Mann, R. J. Oakes and B. Renner, Phys. Rev. {\bf 175}, (1968) 2195.
\bibitem{svz} M. A.~Shifman, A.~I.~Vainstein and V.~I.~Zakharov, Nucl. Phys.
B~{\bf 147}, 385 (1979).
\bibitem{svz2} M. A.~Shifman, A.~I.~Vainstein and V.~I.~Zakharov, Nucl. Phys.
B~{\bf 147}, 448 (1979).
\bibitem{gasser}
  J.~Gasser and H.~Leutwyler,
  Nucl.\ Phys.\ B {\bf 250}, 465 (1985).
\bibitem{bowman}
  P.~O.~Bowman {\it et al.},
  Phys.\ Rev.\ D {\bf 71}, 054507 (2005).
\bibitem{klev} S. P. Klevansky, Rev. Mod. Phys. {\bf 64}, 649 (1992).
\bibitem{wein} S. Weinberg, Phys. Rev. Lett. {\bf 18}, 507 (1967).
\bibitem{avw} P. D. B. Collins, {\it An Introduction to Regge Theory and High-Energy
Physics} (Cambridge University Press, Cambridge, 1977).
\bibitem{afonin:plb}
  S.~S.~Afonin,
  Phys.\ Lett.\ B {\bf 576}, 122 (2003).
\bibitem{Afonin:2014tda}
  S.~S.~Afonin,
  Int.\ J.\ Mod.\ Phys.\ A {\bf 29}, 1450140 (2014).
\bibitem{jaffe}
  R.~L.~Jaffe and A.~Manohar,
  Nucl.\ Phys.\ B {\bf 337}, 509 (1990).
\bibitem{Afonin:2017wjj}
  S.~S.~Afonin,
  Mod.\ Phys.\ Lett.\ A {\bf 32}, 1750179 (2017).
\end{thebibliography}
\end{document}